\documentclass[useAMS,usenatbib,usegraphicx]{mn2e}

\usepackage{amssymb}
\usepackage{threeparttable}
\usepackage{setspace}
\usepackage{float}

\def\reff@jnl#1{{\rm#1\/}}

\def\aj{\reff@jnl{Astron.~J.}}          
\def\araa{\reff@jnl{Ann.~Rev.~Astron.~Astrophys}}   
\def\apj{\reff@jnl{Astrophys.~J.}}      
\def\apjl{\reff@jnl{Astrophys.~J.~Lett.}}  
\def\apjs{\reff@jnl{Astrophys.~J.~Suppl.~Ser.}} 
\def\ao{\reff@jnl{Appl.Opt.}}           
\def\aap{\reff@jnl{Astron.~Astrophys.}} 
\def\aapr{\reff@jnl{Astron.~Astrophys.~Rev.}} 
\def\aaps{\reff@jnl{Astron.~Astrophys.~Suppl.}} 
\def\baas{\reff@jnl{Bull.~Am.~Ast.~Soc.}} 
\def\expast{\reff@jnl{Exp.~Astron.}}    
\def\mnras{\reff@jnl{Mon.~Not.~R.~Ast.~Soc.}} 
\def\pasp{\reff@jnl{Pub.~Astron.~Soc.~Pac.}} 
\def\pra{\reff@jnl{Phys.Rev.A}}         
\def\prb{\reff@jnl{Phys.Rev.B}}         
\def\prc{\reff@jnl{Phys.Rev.C}}         
\def\prd{\reff@jnl{Phys.Rev.D}}         
\def\prl{\reff@jnl{Phys.Rev.Lett}}      
\newcommand\nat{\reff@jnl{Nature}}      
\def\procspie{\reff@jnl{Proc.~SPIE}}    
\def\josa{\reff@jnl{J.~Opt.~Soc.~Am.}}   
                                                                                                                                                                                    
\title[A CO(3-2) survey of luminous infrared galaxies]{A CO(3-2)
  survey of a merging sequence of luminous infrared galaxies}
\author[J. Leech, K. G. Isaak, P. P. Papadopoulos, Y. Gao,
  G. R. Davis]{J. Leech$^{1}$\thanks{E-mail: jxl@astro.ox.ac.uk
    (JL)}, K. G. Isaak$^{2}$, P. P. Papadopoulos$^{3}$,
  Y. Gao$^{4}$, G. R. Davis$^{5}$\\
$^{1}$Astrophysics Group, Department of Physics, University of Oxford, Denys Wilkinson Building, Keble Road, Oxford, OX1 3RH, UK.\\
$^{2}$Cardiff University, School of Physics \& Astronomy, Queens Buildings, The Parade, Cardiff, CF24 3AA, UK.\\
$^{3}$Argelander-Institut f¨ur Astronomie, Auf dem H¨ugel 71, D-53121 Bonn, Germany.\\
$^{4}$Purple Mountain Observatory, Chinese Academy of Science,2 West Beijing Road, Nanjing 210008, China.\\ 
$^{5}$Joint Astronomy Centre, 660 N. A'ohoku Place, Hilo, Hawaii, 96720, USA.\\}
\begin{document}

\date{}

\pagerange{\pageref{firstpage}--\pageref{lastpage}} \pubyear{2008}

\maketitle

\label{firstpage}

\begin{abstract}

  Luminous infrared galaxies ($L_{\rm{IR}}>10^{11} L_{\odot}$) are
  often associated with interacting galactic systems and are thought
  to be powered by merger--induced starbursts and/or dust--enshrouded
  AGN. In such systems, the evolution of the dense, star forming
  molecular gas as a function of merger separation is of particular
  interest. Here, we present observations of the CO(3-2) emission from
  a sample of luminous infrared galaxy mergers that span a range of
  galaxy-galaxy separations. The excitation of the molecular gas is
  studied by examining the CO(3-2)/CO(1-0) line ratio, $r_{31}$, as a
  function of merger extent. We find these line ratios, $r_{31}$, to
  be consistent with kinetic temperatures of $T_k$=(30--50)~K and
  gas densities of $n_{\rm{H}_2}=10^3 \, \rm{cm}^{-3}$. We also find
  weak correlations between $r_{31}$ and both merger progression and
  star formation efficiency ($L_{\rm{fIR}} / L_{\rm{CO(1-0)}}$).
  These correlations show a tendency for gas excitation to increase as
  the merger progresses and the star formation efficiency rises. To
  conclude, we calculate the contributions of the CO(3-2) line to the
  850~$\mu$m fluxes measured with SCUBA, which are seen to be
  significant ($\sim$24\%).

\end{abstract}


\section{Introduction}

 Luminous Infrared Galaxies (LIGs) ($L_{\rm{IR}}>10^{11} L_{\odot}$)
 and Ultra Luminous Infrared Galaxies (ULIGs) ($L_{\rm{IR}}>10^{12}
 L_{\odot}$) are the dominant population for galaxies with bolometric
 luminosities greater than $10^{11} L_{\odot}$ in the local universe
 $(z \lesssim 0.3)$ \citep{1986ApJ...303L..41S}. Observations have
 shown that most are associated with merging/interacting galaxies
 which are rich in molecular gas (see \citet{1996ARA&A..34..749S} for
 an extensive review). Analysis of the original Infrared Astronomical
 Satellite (IRAS) data showed that LIGs are quite rare in the local
 universe but increase significantly with increasing redshift
 \citep{1987ApJ...316L..15H,2005ApJ...622..772C}. Studies of
 \emph{local} luminous infrared galaxies have, therefore, become
 essential for a wider understanding of the role of mergers and
 starbursts both in the local universe and at higher redshifts, where
 such objects are more commonly found.

 Optical studies of local LIGs have shown that many are associated
 with interacting or merging galactic systems
 \citep{1987AJ.....94..831A,1998ApJ...492L.107S,2000ApJ...545..228A,2002ApJS..143..315V}. It
   is now thought that as a galactic merger progresses, gas and dust
   from the parent galaxies dissipate energy through shocks, giving
   rise to an in-fall of gas and dust towards the centre of gravity of
   the interacting system \citep{1996ApJ...471..115B,1996ApJ...464..641M,2006MNRAS.373.1013C}. The
   concentration of cold, dense material then acts to fuel starburst
   activity, or possibly an active galactic nucleus (AGN), or both
   \citep{1998ApJ...508..627K}. The energy produced by the starburst
   or AGN in the optical and ultraviolet is absorbed and then
   re-radiated by dust, leading to an intense infrared luminosity.

 The effect of a galactic merger upon star formation, as well as its
 effect upon the total mass and properties of the molecular and atomic
 gas within the interacting system, has been the subject of several
 observational studies. \citet{1999ApJ...512L..99G} used the CO(1-0)
 spectral line strength as an indicator of total molecular gas mass,
 $M_{\rm{H}_2}$ and plotted this quantity against the projected
 distance of the merging galactic nuclei at optical wavelengths, the
 \emph{projected component separation}, in a merging sample of
 LIGs. The authors found a positive correlation between these two
 quantities. The authors also found an anti-correlation between the
 ratio of infrared luminosity to molecular gas mass,
 $L_{\rm{IR}}/M_{\rm{H}_2}$, a measure of star formation efficiency,
 and the component galaxy separation for their sample. Both
 observations are consistent with a depletion of gas as a huge
 increase in star formation (to around $\sim 100 \, M_\odot \,
 \rm{yr}^{-1}$) occurs. In a survey of literature data,
 \citet{2000MNRAS.318..124G} found strong evidence for gas depletion
 and increased star formation efficiency in a merging galaxy sample
 which included pre--merger and post--merger candidates. The authors
 also found an increase in the central molecular hydrogen surface
 density, as traced by CO(1-0), and a decrease in the fraction of cold
 gas mass as the galactic mergers progressed. This result is
 consistent with large molecular gas inflows, as predicted by
 numerical simulations, the conversion of neutral hydrogen to
 molecular hydrogen, stars and hot gas and molecular hydrogen
 depletion due to ongoing star formation.

  Merging and starburst activity also affects the atomic component of
  the ISM. \citet{1989ApJ...340L..53M} have found that the ratio of
  molecular hydrogen to atomic hydrogen gas mass, $M_{\rm{H}_2}/
  M_{\rm{HI}}$, increases with $L_{\rm{IR}}$. It is currently unclear
  whether this increase is chiefly due to a depletion of HI via
  ejection from the interacting system, depletion of HI due to
  photo-ionisation caused by the formation of young stars or enhanced
  formation of molecular clouds in merger induced shocks. Further
  observations of $M_{\rm{HI}}$ have been attempted to investigate
  whether this depletion of atomic gas is also seen across a merging
  sequence of LIGs \citep{2001A&A...368...64V}, but no
  conclusive answer has yet been derived.

 The evolution of molecular hydrogen gas, which is associated with
 star formation, within merging luminous infrared galaxies is of
 particular interest. Observations of the CO(1-0) rotational
 transition are often used to trace the total molecular hydrogen gas
 mass e.g. \citet{young:scoville}. After $\rm{H}_2$, CO is the most
 abundant molecule in the ISM ($[\rm{CO}/\rm{H}_2] \sim 10^{-4}$), and
 the CO(1-0) transition is typically thermalised in molecular clouds,
 making it a very convenient transition to observe. The low excitation
 energy, $E_{10}/k_B = 5.5$ K and the low critical density
 $n_{\rm{H}_2,\rm{crit}} \sim 410\, \rm{cm}^{-3}$ of the CO(1-0)
 transition make it ideal for tracing the bulk metal--enriched
 $\rm{H}_2$ in galaxies. These same properties, however, make CO(1-0)
 insensitive to the physical conditions of the molecular gas. If we
 require constraints on the density and kinetic temperature, we can
 observe the higher-J CO lines, (i.e. $\rm{CO}_{J+1 - J}$, where
 $J\geq2$) \citep{2003ApJ...588..243Z}.  These transitions will only
 become thermalised and luminous for the denser and warmer gas
 components and it is these dense components which are a direct
 indication of the star formation potential of the interacting
 systems.
 
  The higher order CO(3-2) transition, in particular, is commonly used
  to trace the warmer, denser components of the ISM associated with
  star formation
  \citep{1999A&A...341..256M,2006A&A...460..467B,2009ApJ...693.1736W}
  and constrain the kinetic temperature and density of the molecular
  gas. It is likely that there will be multiple density components of
  molecular gas within the ISM of luminous infrared galaxies, and the
  critical density of CO(3-2) \citep[$n_{\rm{H}_2,\rm{crit}}> 8.4 \times
  10^3 \, \rm{cm}^{-3}$,][]{1995PhDT.......310J} is well matched to
  the density of the \emph{star forming} component. The relative
  ratios of the higher-order CO line intensities depend on both the
  density and the kinetic temperature of the molecular gas
  \citep{1974ApJ...187L..67S,1974ApJ...189..441G}.  By measuring a
  range of CO line ratios one can therefore begin to constrain the
  likely conditions of the molecular interstellar medium. Observing
  the CO(3-2) transition is convenient as it traces mostly the
  starforming molecular gas and has relatively large luminosities
  compared to other commonly used tracers, such as HCN
  ($n_{\rm{H}_2,\rm{crit}}> 10^5 \, \rm{cm}^{-3}$), which have higher
  dipole moments, but lower abundances in the ISM. Measuring the
  CO(3-2) emission is also important since this emission can
  contaminate 850~$\mu$m continuum observations of the dust component
  on the ISM for these galaxies
  \citep{2000ApJ...537..631P}. Correcting this 850~$\mu$m flux for CO
  emission is important when using this flux to determine the dust
  mass and kinetic temperature \citep{2004MNRAS.349.1428S}.

  In this paper, we present observations of the CO(3-2) rotational
  transition to investigate the evolution of the \emph{dense}
  ($n_{\rm{H}_2,\rm{crit}}> 8.4 \times 10^3 \, \rm{cm}^{-3}$)
  molecular gas component as a function of merger extent in a sample
  of LIG mergers. Our aim is to study the evolution of merging
  luminous infrared galaxies by observing a LIG sample at differing
  stages of interaction, and thus obtain a detailed understanding of
  how the starburst evolves as the interaction proceeds. Such
  observations are key to obtaining a \emph{temporal} understanding of
  the evolution of the molecular gas reservoir and how it relates to
  star formation activity as the merging component galaxies become
  increasingly tidally disrupted. Our study builds on work reported by
  \citet{1999ApJ...512L..99G}, in which a sample of luminous galaxies
  was chosen to represent a merging sequence with a range of nuclear
  separations of the merging components. We observe the CO(3-2)
  transition in a subset of this merging sample of 49 LIGs to trace
  the evolution of the warmer, denser molecular gas component as a
  function of merger separation. In particular, we investigate how the
  excitation of the molecular gas changes by examining the evolution
  of the {CO(3-2)/CO(1-0)} line ratio, $r_{31}$, as merging
  progresses. $r_{31}$ will be larger for warmer, denser gas and one
  might expect its value to increase with decreasing nuclear
  separation, as an increasing fraction of the total molecular gas
  becomes concentrated in star forming regions. We then examine how
  the line ratio varies with merging extent, far-infrared luminosity
  ($L_{\rm{fIR}}$, defined in Section \ref{observations}) and star
  formation efficiency ($L_{\rm{fIR}} / L_{\rm{CO(1-0)}}$), and
  compare our sample with two samples with smaller infrared
  luminosities.  Finally we determine contamination arising from the
  CO(3-2) line flux upon the 850~$\mu$m continuum observations of the
  dust component of the ISM for these galaxies. Section
  \ref{observations} outlines the observations that were made,
  including the sample selection criterion, the choice of pointings
  and calibration of the data. The data reduction and results of the
  observations are presented in Section \ref{results} and their
  interpretation and correlations with other data for the sample are
  discussed in Section \ref{interpretation}.  Section
  \ref{conclusions} outlines the conclusions which may be drawn from
  this study.

\section{Sample Selection and Observations} 
\label{observations}

\subsection{The sample}

 Our sample of merging galaxies is a subset of a sample of merging
 galaxies defined by \citet{1999ApJ...512L..99G}, chosen such that the
 sources are observable by the James Clerk Maxwell Telescope
 (JCMT).\footnote{ The JCMT is run by the UK Science and Technology
   Facilities Council, the National Research Council, Canada, and the
   Netherlands Organisation for Scientific Research.}
 \citet{1999ApJ...512L..99G} chose a parent sample of 49 objects to
 include all LIGs with (a) available CO(1-0) and R-band CCD data (b)
 $L_{\rm{IR}}\geq 2 \times 10^{11} L_{\odot}$\footnote{Infrared
   luminosities follow standard IRAS definitions: $L_{\rm{IR}} \equiv
   L(8-1000 \mu \rm{m})=4 \pi D^2_L F_{\rm{IR}} [\rm{L_\odot}]$ and
   $L_{\rm{fIR}} \equiv L(40-500 \mu \rm{m})=4 \pi D^2_L C
   F_{\rm{fIR}} [\rm{L_\odot}]$ where $F_{\rm{IR}} = 1.8 \times
   10^{-14}(13.48 f_{12} +5.16 f_{25} +2.58 f_{60}+f_{100}) \,\rm{W
     m^{-2}}$ and $F_{\rm{fIR}} = 1.26 \times 10^{-14}(2.58 f_{60}
   +f_{100}) \,\rm{W m^{-2}}$
   \citep{1996ARA&A..34..749S,1987PhDT.......114P,1988ApJS...68..151H}.},
 and (c) having nuclear separations ($S_{\rm{sep}}$) greater than
 $2''$ and less than half of the sum of the galactic optical disc
 sizes. The constraint on nuclear separations was set to ensure that
 all genuinely interacting systems were included in the sample, whilst
 excluding difficult to measure, late mergers with separations of
 $S_{\rm{sep}}<2''$, as they are indistinguishable from merged
 single-nucleus galaxies with the usual seeing limit for optical
 observations.\footnote{Since the publication of
   \citet{1999ApJ...512L..99G}, the nuclear separation measurement for
   IRAS 17208-0014 has been improved and the object has been found to
   be a late merger ($S_{\rm{sep}}<2''$) -- we thus exclude it from
   all further sub-samples discussed in this paper.}

 The sample chosen for this study, hereafter referred to as the
 CO(3-2) sample, comprises of 33 objects from the Gao \& Solomon
 sample, chosen to meet the following selection criteria:

\begin{enumerate}

\item A declination such as to be observable from the James Clerk Maxwell Telescope ($\delta > -40^o$).\\

\item A distance such that the redshifted CO(3-2) emission line falls
within the 315-373~GHz tuning range of the 345~GHz receiver, and at least
2 GHz either side of the strong 325~GHz atmospheric water vapour
absorption feature (i.e.  $0 < cz < 17230$~km/s and $21160 < cz <
29310$~km/s).

\end{enumerate}

\begin{figure*}
\begin{center}
\includegraphics [height=7in, angle=-90] {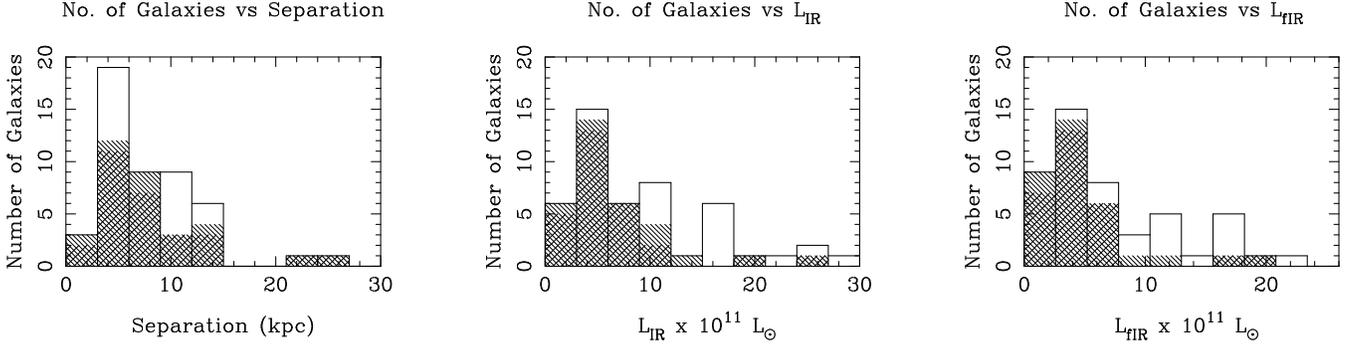}
\caption{Nuclear separation (kpc), infrared luminosity $L_{\rm{IR}}$
  and far infrared luminosity $L_{\rm{fIR}}$ statistics for the
  galaxies in the Gao and Solomon sample (open rectangles), the
  fraction of the sample observable at the JCMT (narrow cross-hatch)
  and the 28 galaxies observed in this study (wide cross-hatch).}
\label{all_hist}
\end{center}
\end{figure*}

 Shown in Figure \ref{all_hist} is the statistical spread of nuclear
 separation, infrared luminosity and far-infrared luminosity, for the
 Gao \& Solomon sample  and the CO(3-2) sample defined here.


\subsection{Choice of pointings}

  The beam width (FWHM) for Receiver B3 on the JCMT is $ 14 ''$ and
  the nuclear separations for our sample lie between $2.5''$ and
  $40.6''$, which made it necessary to choose suitable pointings
  carefully so as to capture all of the CO(3-2) emission.  The choices
  of pointing centre were made based on nuclear separation, which in
  turn were derived from either images taken at R-band from
  \citet{Sanders:Rband}, 645 nm (band 103aE) images from the STScI
  Digitized Sky Survey (DSS) \citep{1994Ap&SS.217...27K} or with WFPC2
  images taken with the Hubble Space Telescope.  Where available,
  high-resolution CO(1-0) interferometry maps were also used to
  confirm the choice of pointings
  \citep{2001AJ....122..140W,1999A&A...346..663C,1997ApJ...475L.103L}.
  Three different categories of pointings/observations were
  identified:

\begin{enumerate}

\item Objects with a projected nuclear separation of $\lesssim 7''$
 (less than half of beam size): single pointing sufficient to capture
 all of the CO(3-2) emission. 16 objects of our sample were observed
 with a single pointing. 

\item Objects with a projected nuclear separation of $\sim 14''$ (of
order of the beam size): two pointings used, centred on the nuclei
of the merging galaxies. 7 objects of our sample were observed with a dual pointing.

\item Objects with a projected nuclear separation $> 14''$: multiple
  pointings, separated by $1/2$ beam ($7''$) were used to realise a
  fully sampled strip along the axis joining the two component
  nuclei. 5 objects of our sample were observed with multiple
  pointings (see Figure \ref{spec1}).

\end{enumerate}

\begin{table*}
\begin{minipage}{140mm} 
\begin{spacing}{1}
\caption{Observational parameters for the LIGs observed in this study.}
\begin{threeparttable}
\centering
\scriptsize
\begin{tabular}{llccccccc} \hline\hline
Object    & \multicolumn{2}{c}{Map Centre (J2000)}& cz \tnote{a} & Offsets  & $I_{\rm{CO}}(3-2)$ & $ \sigma I_{\rm{CO}}(3-2)$ & Effective        \\
          &  R.A.           & Dec.                &(km/s)        & $('','')$& (K km/s)           & (K km/s)                 & linewidth (km/s) \tnote{b} \\\hline

00057+4021 &  00:08:21.0   &  +40:37:56           & 13516 & (0,0)       & 9.5                &  0.5      & 221    \\ 
Arp 236    &  01:07:47.0   &  -17:30:24           &  6016 & (0,0)       & 107.8              &  2.1      & 301    \\
	   &	           &	                  &       & (6,-2)      & 98.5               &  2.1      & 295    \\
	   &	           &	                  &       & (-7,1)      & 51.8               &  2.1      & 312    \\
01077-1707 &  01:10:08.2   & -16:51:11            & 10540 & (0,0)       & $<$3.8             &  1.9      &  -     \\
	   &	           &	                  &       & (-21,0)     & $<$3.0             &  1.5      &   -    \\
	   &	           &	                  &       & (-14,0)     & $<$2.4             &  1.2      &        \\
	   &	           &	                  &       & (-7,0)      & $<$4.4             &  2.2      &        \\ 
	   &	           &	                  &       & (7,0)       & 15.3               &  2.3      &  230   \\
	   &	           &	                  &       & (14,0)      & 24.6               &  2.4      &  200   \\
	   &	           &	                  &       & (21,0)      & 15.8               &  2.1      &  195   \\
III Zw 35  &  01:44:30.5   & +17:06:08            & 8215  & (0,0)       & 18.5               &  2.2      &  262   \\
Mrk 1027   &  02:14:05.6   & +05:10:27.7          & 8913  & (0,0)       & 28.5               &  1.2      &  283   \\
02483+4302 &  02:51:36     & +43:15:11            & 15419 & (0,0)       & 5.74               &  0.7      &  163   \\
UGC 2369   &  02:54:01.8   & +14:58:14            &  9354 & (0,0)       & 20.5               &  2.3      &  302   \\
           &	           &	                  &       & (-6,20)     &  $<$5.8            &  2.9      &   -    \\
03359+1523 &  03:38:46.9   &  +15:32:55           & 10600 & (0,0)       &  12.3              &  2.5      &  181   \\
           &	           &	                  &       & (-12,2)     &  $<$4.4            &  2.2      &   -    \\
04232+1436 &  04:26:04.8   &  +14:43:37           & 23855 & (0,0)       &  20.0              &  2.4      &  639   \\
Arp 55     &  09 15 54.9   &  +44 19 54.4         & 11773 & (0,0)       &  28.1              &  2.6      &  258   \\
           &               &                      &       & (10,7)      &  18.3              &  3.4      &  294   \\
10039-3338 &  10:06:04.5   &  -33:53:03           & 10235 & (0,0)       &  10.9              &  0.4      & 146    \\
10190+1322 &  10:21:42.6   &  +13:06:56           & 22987 & (0,0)       &  10.3              &  1.1      & 291    \\     
10565+2448 &  10:59:18.1   &  +24:32:34           & 12926 & (0,0)       & 28.4               &  0.9      & 187    \\    
Arp 299    &  11:28:33.6   &  +58:33:44           &  3115 & (0,0)       & 79.1               &  2.9      & 261    \\          
	   &	           &	                  &       & (-7,1)      & 52.4               &  2.7      & 192    \\       
	   &	           &	                  &       & (-14,2)     & 66.8               &  2.7      & 175    \\      
	   &	           &	                  &       & (-19,4)     & 84.3               &  2.7      & 167    \\      
13001-2339 &  13:02:52.1   &  -23:55:19           & 6446  & (0,0)       & 43.8               &  1.5      & 479    \\     
Arp 238    &  13:15:30.2   &  +62:07:45           & 9453  & (0,0)       & $<$5.0             &  2.5      &  -     \\       
           &	           &	                  &       & (31,-17)    & 16.3               &  2.7      & 240    \\ 
NGC 5256   &  13:38:17.9   &  +48:16:41           &  8239 & (0,0)       & 18.4               &  2.5      & 213    \\ 
 	   &	           &	                  &       & (-7,-9)     & 29.0               &  4.2      & 325    \\  
Mrk 673    &  14:17:21.0   &  +26:51:28           & 10980 & (0,0)       & 18.9               &  2.9      & 312    \\                
14348-1447 &  14 37 38.27  & -15 00 22.2          & 24677 & (0,0)       & 19.4               &  2.0      & 240    \\     
Arp 302    &  14:57:00.7   &  +24:37:03           & 10166 & (0,0)       & 48.1               &  3.0      & 502    \\            
           &               &                      &       & (0,8)       & 26.4               &  1.7      & 216    \\            
           &	           &	                  &       & (-1,-7)     & 26.0               &  2.6      & 255    \\            
           &               &                      &       & (-2,-14)    & $<5.2$             &  2.6      &  -     \\           
           &               &                      &       & (-3,-21)    & $<4.4$             &  2.2      &  -     \\            
           &               &                      &       & (-3,-28)    & $<5.2$             &  2.6      &  -     \\           
           &               &                      &       & (-4,-37)    & 8.79               &  1.1      & 74     \\           
Mrk 848    & 15:18:06.2    &  +42:44:42           &12043  & (0,0)       & 19.9               &  1.1      & 121    \\ 
NGC 6090   & 16:11:40.7    &  +52:27:25           &8754   & (0,0)       & 56.4               &  2.2      & 141    \\              
17132+5313 & 17 14 20.38   &  +53 10 30.4         & 15290 & (0,0)       & 12.5               &  2.5      & 221    \\       
NGC 6670   &  18:33:37.7   &  +59:53:24           &  8684 & (0,0)       & 13.9               &  1.2      & 151    \\                
	   &	           &	                  &       & (-41,-15)   & $<$3.8             &  1.9      &  -     \\          
           &	           &	                  &       & (-35,-11)   &  11.4              &  1.9      & 166    \\         
           &               &	                  &       & (-28,-9)    & $<$4.0             &  2.0      &  -     \\          
           &               &	                  &       & (-21,-6)    & 19.2               &  2.5      & 245    \\         
           &	           &                      &       & (-14,-3)    & 12.1               &  1.8      & 174    \\        
	   &        	   &	                  &       & (-7,-2)     & 5.7                &  1.0      & 100    \\   
19297-0406 &  19:32:21.3   &  -03:59:54           & 25674 & (0,0)       & $<$1.8             & 0.9       & -      \\     
20010-2352 &  20:04:03.5   &  -23:43:54           & 15249 & (0,0)       & 16.4               & 1.4       & 344    \\           
II Zw 96   &  20:57:23.7   &  +17:07:44           & 10900 & (0,0)       & 13.7               & 1.4       & 176    \\  
           &	           &	                  &       & (8,-9)      & 17.9               & 1.4       & 180    \\            
NGC 7592   &  23:18:22.6   &  -04:24:58           &  7328 & (0,0)       & $<$5.2             & 2.6       &  -     \\        
           &	           &	                  &       & (-12,2)      & $<$6.2            & 3.1       &  -     \\  \hline
\end{tabular}
\begin{tablenotes}
\item[a] {Optical redshift definition, values obtained from NASA/IPAC Extragalactic Database (NED) {\tt http://nedwww.ipac.caltech.edu/}}.
\item[b] {Effective linewidth $\equiv T_{\rm{mb,peak}}/ I_{\rm{CO}}(3-2)$, where $T_{\rm{mb,peak}}$ is the peak main--beam brightness temperature of the line.}
\end{tablenotes}
\label{obs_table}
\end{threeparttable}
\end{spacing}
\end{minipage}
\end{table*}

\subsection{CO(3-2) observations} 

Observations of the CO(3-2) line ($\nu_{rest}=345.796$~GHz) were made
using the dual-channel (orthogonal polarisation), 345~GHz receiver
(RxB3) on the JCMT. The DAS (Digital Auto-correlation
Spectrometer) backend was used in one of two bandwidth modes,
depending on the expected width of the CO(3-2) spectral line given the
CO(1-0) linewidths. The single-channel, wideband mode, in which the
two 900~MHz subbands are concatenated to realise a useable contiguous
bandwidth of 1800~MHz ($\Delta v \sim 1560$~km/s), was used for
expected line widths of $\Delta_{\rm{FWHM}}> 500$km/s; the more sensitive
dual-channel standard mode, with a single subband of 900~MHz ($\Delta
v \sim 780$ km/s) was used where lines of $\Delta v_{\rm{FWHM}}< 500$km/s
were anticipated. The wideband mode was only used where the velocity
covered by standard mode was not sufficient to provide line--free
regions for baseline removal.

 Rapid beam--switching at a secondary mirror chop frequency of 1-2 Hz
and beam throws of $60-90''$ in azimuth were used to ensure good
atmospheric cancellation.  Hot/cold load calibrations were performed
every 20 to 30 minutes to achieve consistent calibration to the
$T^*_A$ antenna temperature scale. The telescope pointing was checked
every 2--3 hours using local spectral line pointing sources, and any
offset corrections applied to the telescope pointing model, resulting
in an RMS pointing accuracy of 1--2$''$. The science data taken for
each pointing consisted of spectra from each RxB3 polarisation
channel, usually integrated for a period of around 10
minutes. Integration times of longer than 10 minutes were achieved by
coadding several $\sim10$ minute spectra while checking the data for
instrumental drifts and baseline problems.

 Spectral line standards such as OMC1 were observed to check amplitude
 calibration and quantify calibration uncertainties, as well as to
 monitor the overall performance of the telescope. Mars and Uranus,
 when observable, were observed to establish main-beam efficiencies.

 A total of 28 of the 33 sources in the CO(3-2) sample were observed
 over the period between October, 2000 - July, 2003. The atmospheric
 opacity at 225 GHz measured at the CSO, was typically $\tau_{225
   \rm{GHz}} \sim 0.1$ giving system temperatures of around 600~K in
 the receiver tuning range used for the observations.  Of our 28
 observed sources , 16 / 7 / 5 were observed with single / two /
 multiple pointings. Figure~\ref{all_hist} shows a comparison of the
 statistical spread of nuclear separations, infrared luminosities and
 far-infrared luminosities for the objects actually observed.

\section{Data reduction and results}
\label{results}

 All data were reduced using the JCMT spectral line reduction package
 SPECX \citep{specx}. Individual spectra were inspected, with any
 scans showing significant baseline structure/ripple rejected ($\sim
 5$\% of total). The spectra were then coadded, linear baselines
 fitted and removed and then the spectra were binned to 26~km/s
 (30~MHz). The $I_{\rm{CO}}(3-2)$ values were determined by directly
 integrating under the extent of the spectral line.

  Planetary observations of Mars and Uranus were used to estimate
$\eta_{\rm{mb}}$ and calibrate the data directly from the
 sky-corrected antenna temperature, ${{T^{*}_A}}$ to the main-beam
 brightness temperature $T_{\rm{mb}}$ via

\begin{equation}
\eta_{\rm{mb}} \equiv B_{\rm{eff}}/F_{\rm{eff}} = {{T^{*}_A}\over{T_{\rm{mb}}}}. 
\label{eta_mb}
\end{equation}

 \noindent Here, $B_{\rm{eff}}$, the beam efficiency, and
 $F_{\rm{eff}}$, the forward efficiency follow the definitions given
 in \citet{1988dli..conf...53D}. Values of $\eta_{\rm{mb}}\sim 0.55$
 were typically measured using this technique. Where no planetary
 observations were available, a canonical value for RxB3 on the JCMT
 of $\eta_{\rm{mb}}=0.55$ was adopted, which reflects the mean value
 obtained and published on the JCMT web pages for the observing
 period\footnote{\tt \scriptsize
   http://www.jach.hawaii.edu/JCMT/spectral\_line/Standards/beameff.html}. The
 observed spread for $\eta_{\rm{mb}}$ over the period of observing was
 $\sim 20$\%, consistent with that typically seen for RxB3 on the
 JCMT.

 The uncertainty in $\eta_{\rm{mb}}$ introduces a systematic
 calibration uncertainty of $\pm$ 20\%. In addition there are
 random noise contributions arising from (a) noise across each
 spectral channel (b) determining the integrated spectral line and (c)
 fitting the overall subtracted baseline level. These errors can be
 combined, following \citet{Gao:phd} (see Appendix \ref{errors}), to
 give an overall error, $\sigma I_{\rm{CO}}(3-2)$, in the
 velocity-integrated main-beam brightness temperature (the integrated
 line intensity), $I_{\rm{CO}}(3-2)$.

 The efficiency-corrected spectra together with the pointing centres
 superimposed on a Digital Sky Survey optical image, are shown in
 Figure \ref{spec1} at the end of the paper. The properties of the
 spectra for each pointing, together with the random errors, $\sigma
 I$ are summarised in Table \ref{obs_table}. Where no line was
 detected, an upper limit for $I_{\rm{CO}}(3-2)$ of $2 \sigma$ is
 given, assuming the detected line would have had a full width to zero
 intensity of $\Delta W = 400 $~km/s. The observed spectra shown in
 Figure \ref{spec1} show a wide range of effective linewidths (from 74
 to to 639~km/s) and often show double peaked velocity profiles. These
 profiles can arise when a single telescope beam picks up CO(3-2)
 emission from each of the interacting nuclear components or when
 there is rotation of the molecular gas around one of the nuclear
 components within the beam.

\begin{table*}
\begin{minipage}{140mm} 
\begin{spacing}{1}
\caption{Component separations, infrared and CO luminosities and line ratios for our LIG sample.}
\begin{threeparttable}
\centering
\scriptsize
\begin{tabular}{llcccccccc} \hline\hline
Object    &   \multicolumn{2}{c}{Separation}& $L_{\rm{IR}}$ & $L_{\rm{fIR}}$ & $I_{\rm{CO}}(3-2)$ & $\sigma I_{\rm{CO}}(3-2)$ & $L_{\rm{CO}}(3-2)$ & $L_{\rm{CO}}(1-0)$& $r_{31}$ \\
          &  (kpc)          & ($''$)   & $10^{11} L_{\odot}$ & $10^{11} L_{\odot}$ & (K km/s, Total)       & (K km/s)     &  ($10^9 L_l $)\tnote{a} & ($10^9 L_l$)\tnote{a} &\\\hline 
00057+4021 & 2.1  & 2.5  &   4.3 & 3.31 & 9.5  & 0.5 & 1.65 & 3.8 & 0.43 \\   
Arp 236    & 6.1  & 16   &  4.71 & 3.19 & 152  & 3.0 & 5.21 & 10.67 & 0.49 \\ 
01077-1707 & 22.4 & 35   &  4.33 & 2.79 & 24.6 & 2.4 & 2.49 & 8.5 & 0.29 \\ 
III Zw 35  & 4.1  & 8.2  &  4.15 & 3.58 & 18.5 & 2.2 &  1.21 & 2.09 & 0.58 \\   
Mrk 1027   & 5    & 9    & 2.49  & 1.69 & 63.8 & 2.6 & 4.79 & 5.35 & 0.89 \\ 
02483+4302 & 3.6  & 3.8  & 6.21  & 4.54 & 5.74 & 0.7 & 1.30 & 2.88 & 0.45 \\   
UGC 2369   & 13.1 & 22.5 & 3.95  & 2.66 & 20.5 &  2.3 & 1.69 & 7.25 &   0.23 \\ 
03359+1523 & 6.5  & 10   & 3.34  & 2.42 & 12.3 &  2.5 & 1.30 & 6.9  &   0.19 \\ 
04232+1436 & 6.3  & 4.6  & 11.15 & 7.49 & 20.0 &  2.4 & 10.3 & 9.0 & 1.15 \\     
Arp 55     & 10.7 & 15   & 4.66  & 3.39 & 45.1 &  3.8 & 5.74 & 12.07 & 0.48 \\ 
10039-3338 & 3.1  & 5    & 4.54  & 3.94 & 15.3 & 0.6 & 1.46 & 2.93  & 0.50 \\ 
10190+1322 & 6.6  & 5    & 11.34 & 7.13 &  10.3 &  1.1 & 4.86 & 8.2   & 0.59 \\ 
10565+2448 & 6.2  & 8    & 9.58  & 7.59 & 28.4  &  0.9 & 4.31 & 5.76 & 0.75 \\  
Arp 299    & 4.5  & 25   & 6.36  & 4.74 & 183 & 2.8 & 1.64 & 3.0 & 0.55 \\  
13001-2339 & 2.6  & 6.5  & 2.45  & 2.0  & 43.8 &  1.5 & 1.67 & 2.55 & 0.66 \\ 
Arp 238    & 12.1 & 21   & 5.17  & 3.8  & 16.3 &  2.7 & 1.34 & 4.24 & 0.32 \\ 
NGC 5256   & 4.8  & 9.5  & 3.05  & 2.02 & 39.6 &  4.2 & 2.46 & 5.66 & 0.43 \\ 
Mrk 673    & 4.0  & 6    & 2.2   & 1.51 & 18.9 &  2.8 & 2.04 & 4.0 & 0.51  \\  
14348-1447 & 5.7  & 4.0  & 20.43 & 16.28& 19.4 & 2.0 &10.55 & 14.0 & 0.75 \\ 
Arp 302    & 25.8 & 41.6 & 4.1   & 3.59 & 59.0 & 2.4 & 5.59 & 19.7 & 0.28 \\ 
Mrk 848    & 4.8  & 6.5  & 7.16  & 5.29 & 19.9 & 1.1 & 2.62 & 7.03 & 0.37 \\ 
NGC 6090   & 3.5  & 6.5  & 2.98  & 1.94 & 56.4 &  2.2 & 3.96 & 5.0 & 0.79  \\ 
17132+5313 & 6.4  & 6.5  & 7.67  & 5.73 & 12.5 &  2.5 & 2.64 & 7.24 & 0.37   \\  
NGC 6670   & 14.6 & 26.5 & 3.84  & 2.73 & 44.6 & 3.4  & 3.26 & 11.6 & 0.28 \\ 
19297-0406 & 11.5 & 8    & 24.59 & 19.28& $<$1.8 &  1.1 &$<$1.1 & 9.43 & $<$0.11\\ 
20010-2352 & 8.1  & 8.9  & 4.69  & 3.16 &  16.4 &  1.4 & 3.47 & 6.38 & 0.54 \\ 
II Zw 96   & 7.4  & 11   & 7.57  & 6.36 & 27.2 &  1.7  &3.08 & 6.04 & 0.51 \\  
NGC 7592   & 5.3  & 11.5 & 2.43  & 1.68 &$<$6.2 &  3.1 & $<$0.31 & 4.55 & $<$0.07 \\ 
\bf Mean   &\bf 7.15 & \bf 11.5 & \bf 5.2&\bf 4.03&\bf 37.8  &\bf 2.0 &	  \bf 3.2 & \bf 6.5& \bf 0.47 \\\hline 
\end{tabular}
\begin{tablenotes}
\item[a] Calculated using equation (\ref{e1_a}), $H_0 = 75$ km/s $\rm{Mpc}^{-1}$,  $q_0$=0.5, $L_l = \rm{K\, km\,s^{-1}\, pc^2}$.
\end{tablenotes}
\label{derived_obs_table}
\end{threeparttable}
\end{spacing}
\end{minipage}
\end{table*}

\subsection{Luminosities and line ratios}

  Having determined $I_{\rm{CO}}(3-2)$ for each pointing for each
  object, we proceeded to calculate the total CO(3-2) line luminosity
  and the CO(3-2)/CO(1-0) line luminosity ratio for each object as
  follows. For galaxies with a projected nuclear separation of
  $\lesssim 7''$, it was assumed that the angular size of the CO(3-2)
  emitting region was small compared to the $14''$ beam size when
  determining the overall CO(3-2) luminosity. Following
  \citet{1999ApJ...512L..99G}, the CO(3-2) line luminosities,
  $L_{\rm{CO}(3-2)}$, were calculated by following the definition of
  CO(1-0) luminosity given by \citet{1997ApJ...478..144S},

\begin{eqnarray}
\lefteqn{L_{\rm{CO}(3-2)}={T_B}\Delta V \Omega_{s} D^2_A}\nonumber\\
& & =23.5\Omega_{s\star b}D^2_L I_{\rm{CO}(3-2)} (1+z)^{-3} \, \, \,[\rm{K\, km\, s^{-1} \rm{pc}^2}]
\label{e1_a}
\end{eqnarray}

 \noindent where $T_B$ is the peak line brightness temperature,
 $\Delta V$ is the effective line width, $I_{\rm{CO}(3-2)} \equiv \int
 T_{\rm{mb}}dV$, $T_{\rm{mb}}$ is the main beam brightness
 temperature, $D_L$ is the luminosity distance and $\Omega_{s\star b}$
 is the solid angle ($\rm{arcsec}^2$) of the source convolved with the
 beam. A value of $\Omega_{s\star b} = \Omega_b =
 1.133\theta^2_{\rm{fwhm}} = 222\, \rm{arcsec}^2$ was used for objects
 observed with a single pointing, following the above assumption that
 the CO(3-2) emitting region was small compared to the beam. For
 objects observed with multiple pointings the same value was used
 after combining the individual fluxes from each pointing,
 $I_{\rm{CO}(3-2)}$, to calculate a total flux, $I_{\rm{CO}(3-2,
   \rm{Total})}$. Upon examining SCUBA $850 \mu m$ continuum maps for
 Mrk~1027 after we made our CO(3-2) observations we found that the dust
 emission from the source appeared extended ($6''$ by $12''$) and that
 centre of this emission was offset by 3.5$''$ relative to
 our CO(3-2) (0,0) pointing. We thus took $I_{\rm{CO}(3-2,
   \rm{Total})} = 1.6*1.4 I(0,0)$, with the first factor correcting for
 beam dilution and the second factor correcting for the pointing
 offset. For IRAS~10039-3338 the flux was inadvertedly measured at
 3.5$''$ from centre of optical/ IR emission, so we corrected this
 mispointing by using $I_{\rm{CO}(3-2, \rm{Total})}= 1.4 I(0,0)$,
 which assumes the emission is small compared to the beam.
  
  For galaxies observed by more than one pointing it was necessary to
  combine the fluxes to determine the total line luminosity. For
  galaxies with a projected nuclear separation $\sim 14 ''$ observed
  with two pointings, the fluxes were combined as a weighted sum using
  the method presented in Appendix \ref{multiple}. For objects with
  nuclear separation greater than $14 ''$, the fluxes from each
  pointing were combined by assuming a plausible underlying brightness
  distribution based on the object morphology. For IRAS~01077-1707 the
  emission appeared consistent with a single compact source at
  $(14,0)$, so we estimated the total flux via $I_{\rm{CO}(3-2,
    \rm{Total})} = I(14,0)$. For Arp~236, the emission of the source
  appeared to be consistent with a source equal to the beamsize so we
  took $I_{\rm{CO}(3-2, \rm{Total})}=\sqrt{2} I_{\rm{CO}}(3-2) (0,0)$.
  For Arp~299 we combined the fluxes via $I_{\rm{CO}(3-2, \rm{Total})}
  = (I(0,0) + I(-19,4) + I(-7,1) + I(-14,2))/2$, consistent with an
  interferometric CO(1-0) map appearing in
  \citet{1999AA...346..663C}. For Arp~302 we assumed a uniform
  underlying distribution across the pointings at (0,0), (0,8) and
  (-1,-7) and combined these separately with the flux at (-4,-37),
  $I_{\rm{CO}(3-2, \rm{Total})} = (I(0,0) + I(0,8) + I(-1,-7))/2 +
  I(-4,-37)$. For NGC~6670, the emission appeared to be strongest at
  pointings (0,0), (-21,-6) and (-35,-11), so we estimated the total
  flux via $I_{\rm{CO}(3-2, \rm{Total})} = I(0,0) + I(-21,-6) +
  I(-35,-11)$, i.e. assuming that most of the emission is
  concentrated in these directions.  The derived values for the total
  CO(3-2) luminosity $L_{\rm{CO(3-2)}}$ and the CO(3-2)/CO(1-0) line
  luminosity ratios, $r_{31}$ are presented in Table
  \ref{derived_obs_table}.

  The full physical implications of line ratio measurements for
  molecular gas conditions require non-LTE photon transport
  models. These can be solved numerically, accounting for the
  excitation between several energy levels of the molecular species
  present.  A popular approach is the Large Velocity Gradient method
  (LVG) \citep{1974ApJ...187L..67S,1974ApJ...189..441G} which
  assumes that large-scale systematic velocity gradients, rather than
  thermal motions, dominate the observed line widths. This assumption
  simplifies the photon transport problem, since there is essentially
  no overlap in line emission between distant parts of the cloud. For
  the mean value of $\overline{r_{31}}=0.47$ for the LIGs in our
  sample, we performed an LVG fit assuming a
  $^{12}$CO(1-0)/$^{13}$CO(1-0) line ratio, $R_{10}$, of $R_{10}=$
  15--20 measured typically for such galaxies \citep[e.g.][]{
  1992A&A...264...55C,1998ApJ...492..521P}. This LVG
  fit yielded typical conditions of kinetic temperature $T_k$=(30--50)
  K and a $\rm{H}_2$ number density of $n=10^3 \rm{cm}^{-3}$ with a CO
  abundance per velocity gradient $\Lambda \equiv
  [\rm{CO/H_2}]/(dV/dR)$ of $10^{-5}$ $(\rm{km \, s^{-1}}\,
  \rm{pc}^{-1})^{-1}$, though good solutions exist also for a much
  warmer ($T_k \sim$ 80--110~K) and diffuse $(n \sim 3 \times 10^2
  \rm{cm}^{-3})$ gas phase. This degeneracy reflects both the lack of
  constraints that could be set by observing a larger number of
  molecular lines (especially high-J CO transitions, $^{13}\rm{CO}$
  lines, and high density tracers such as HCN transitions), and the
  presence of a diffuse and warm gas phase along with a denser and
  cooler one that usually exists in such systems \citep[e.g.][]{1995A&A...300..369A}.  The hereby observed CO(3-2) line is
  the highest-J CO transition to have significant contributions from
  the former while tracing mostly the later phase
  \citep{2007ApJ...668..815P}. We intend to combine our CO(3-2) data
  with HCN observations for the same galaxy sample to further
  constrain the molecular gas conditions. We also intend to conduct a
  more detailed analysis of several of the LIGs in our sample, as part
  of a multi-J CO and HCN line survey of such systems \citep{2007ApJ...668..815P}.

\subsection{Statistical properties of the sample}

\begin{figure}\begin{center}
\includegraphics [width=2.5in,angle=-90] {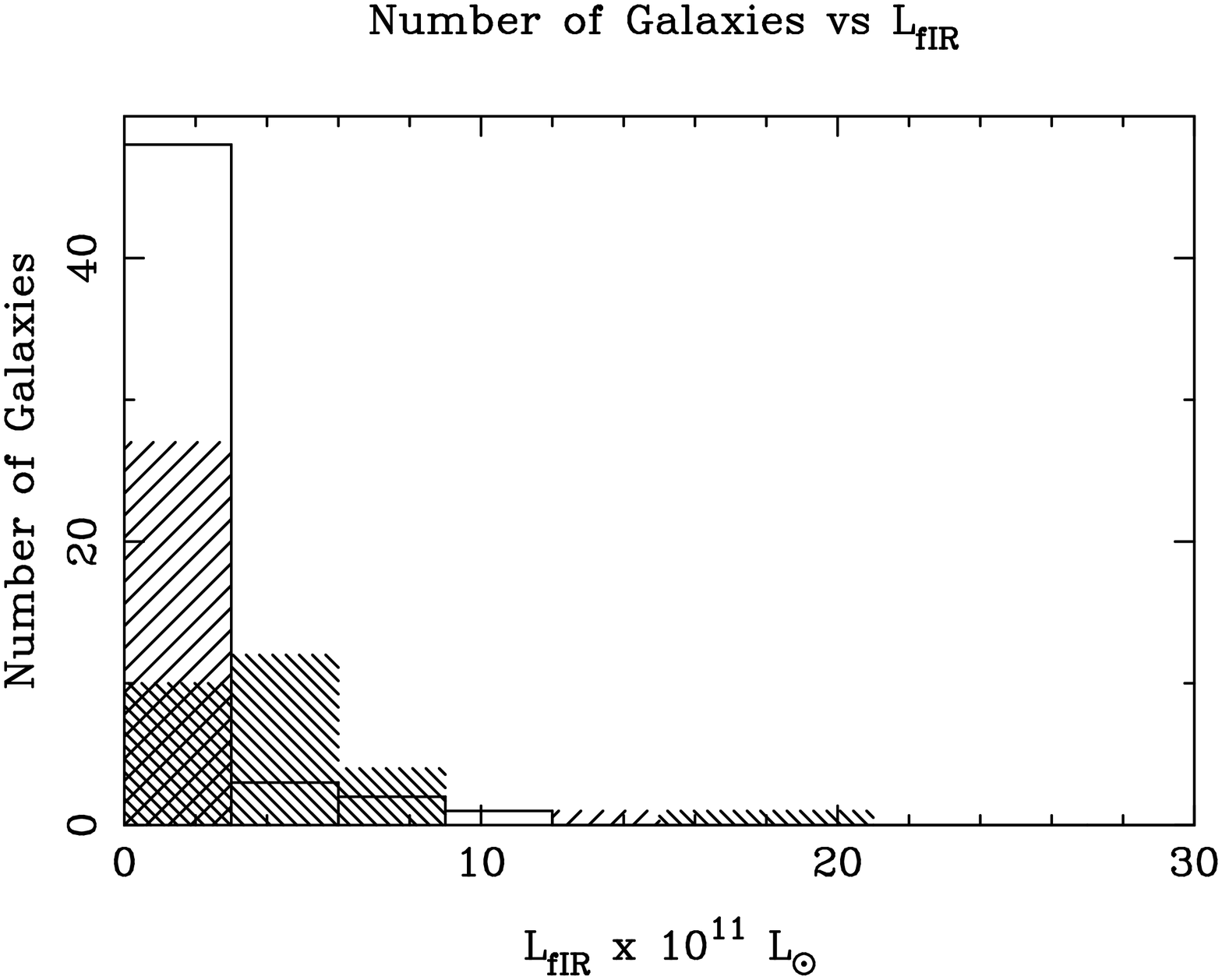}
\caption{ A comparison of the $L_{\rm{fIR}}$ for the SLUGS sample
  (open rectangles), our sample (narrow cross-hatch) and the
  Mauersberger sample (wide cross-hatch). }
\label{lfir_samples} 
\end{center}
\end{figure}

\begin{figure}\begin{center}
\includegraphics [width=2.5in,angle=-90] {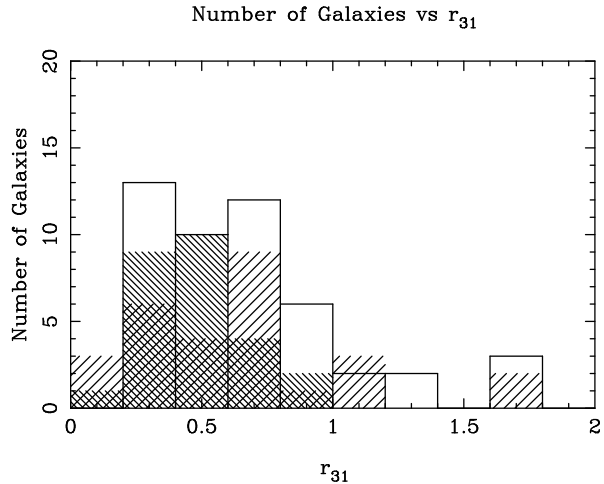}
\caption{ A comparison of the $r_{31}$ for the SLUGS sample
  (open rectangles), our sample (narrow cross-hatch) and the
  Mauersberger sample (wide cross-hatch). }
\label{rat_samples} 
\end{center}
\end{figure}

  It is instructive to compare the CO(3-2) observations of the LIG
  sample here with the CO(3-2) observations of a sample of 29 local
  galaxies reported by \citet{1999A&A...341..256M} (Figures
  \ref{lfir_samples} and \ref{rat_samples}). The authors selected
  galaxies which exhibit strong $I_{\rm{CO}}(1-0)$ and
  $I_{\rm{CO}}(2-1)$ intensities, though mostly due to their proximity
  rather than high intrinsic CO(1-0) or CO(2-1) luminosity. Their
  sample contained 27 non-infrared luminous galaxies with $ 0.9 \times
  10^9 L_\odot < L_{\rm{IR}} < 96\times 10^9 L_\odot$ and 2 LIGs,
  NGC~2146 ($L_{\rm{IR}} =1.1\times 10^{11} L_\odot $) and Arp~220
  ($L_{\rm{IR}} =1.3\times 10^{12} L_\odot $). The majority (26) of
  the galaxies were isolated spirals of type SAB, SA or SB. The
  CO(3-2)/CO(1-0) line ratio for most of the galaxies was between 0.2
  and 0.7 with only 4 objects exhibiting line ratios greater than
  unity. Thus, most of our high infrared luminosity sample have a
  similar range of CO(3-2)/CO(1-0) line ratios to the Mauersberger
  sample i.e. a selection of nearby galaxies with moderate infrared
  luminosity.

  We can also compare our CO(3-2) observations with those of a
 subsample of the SCUBA Local Universe Galaxy Survey (SLUGS) presented
 by \citet{2003ApJ...588..771Y}. The 60 galaxies chosen were a
 near-complete, flux-limited sub-sample $S_{60 \mu m} > 5.24\, \rm{Jy}$
 sample with $L_{\rm{fIR}}$ predominantly less than $10^{11} L_\odot$. Their
 sub-sample is therefore on average considerably less IR-luminous than
 our sample CO(3-2) (Figure \ref{lfir_samples}), yet we see from Figure
 \ref{rat_samples} that a similar spread of line ratios is
 observed with the majority exhibiting $r_{31} < 1$.



%

\section{Discussion}
\label{interpretation}

\subsection{Evolution across the merging sequence}
  
 Since our sample represents a merging sequence of galaxies, we can
 now examine how the dense molecular gas component, traced by the
 CO(3-2) transition, evolves as merging progresses. In particular we
 will determine the correlation between the CO(3-2)/CO(1-0) line
 ratio,  $r_{31}$, and nuclear component separation,  $L_{\rm{fIR}}$
 and  $L_{\rm{fIR}} / L_{\rm{CO(1-0)}}$, a commonly used measure
 of star formation efficiency.

\subsubsection*{Trend with nuclear separation}

\begin{figure}\begin{center}
\includegraphics [width=2.2in,angle=-90] {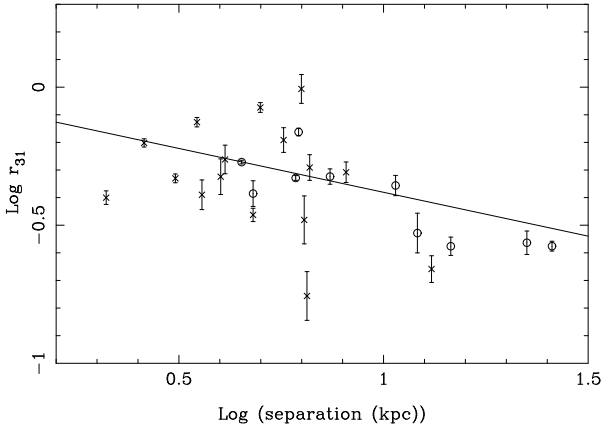}
\caption{ $\log(r_{31})$ vs. the logarithm of the nuclear separation
  for the detected objects in the LIG sample. Crosses indicate objects
  observed with a single pointing, whilst open circles denote objects
  observed with multiple pointings. The error bars reflect $1-\sigma$
  errors in the CO(3-2) line intensities, and exclude $\sim20\%$
  absolute calibration errors.  The best fit line, weighted by the
  random errors, is shown for illustration, and the Pearson
  correlation coefficient is -0.52. }
\label{log_sep_vs_log_rat} 
\end{center}
\end{figure}

\begin{figure}\begin{center}
\includegraphics [width=2.2in,angle=-90] {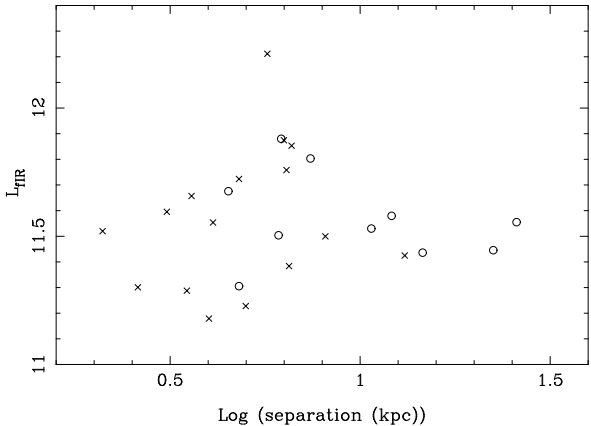}
\caption{A plot of $\log L_{\rm{fIR}}$ vs.  the logarithm
  of the nuclear separation for the detected objects in the LIG
  sample, with symbols defined as in Figure
  \ref{log_sep_vs_log_rat}. The Pearson correlation coefficient is
  0.06.}
\label{log_fIR_vs_log_sep} 
\end{center}
\end{figure}

 Shown in Figure \ref{log_sep_vs_log_rat} is a plot of $\log(r_{31})$
 vs. the logarithm of the nuclear separation ($\log(\rm{sep\,,
   kpc})$), for which we see a weak anticorrelation ($r=-0.52$). We
 observe no correlation between $\log (L_{\rm{fIR}})$ and the
 logarithm of the nuclear separation (Figure
 \ref{log_fIR_vs_log_sep}). This lack of correlation is unsurprising,
 given that $\log (L_{\rm{fIR}})$ is expected to depend on a number of
 different factors (e.g. the total extent of the star forming regions,
 dynamical mass of overall merger), which in turn are not simply, if
 at all, related to the progression of the merger.  In contrast, by
 looking at the CO(3-2)/CO(1-0) line {\it ratio} we are in effect
 independent of the overall mass of the merger system.  If we assume,
 therefore that $r_{31}$ is a measure of the extent of excitation of
 the molecular gas present in the merging galaxies, then the weak
 anti-correlation ($r=-0.52$) is consistent with the gas excitation
 increasing as the merger progresses.

\subsubsection*{Correlations between line ratio, $L_{fIR}$ and star formation 
efficiency} 

\begin{figure}\begin{center}
\includegraphics [width=2.2in,angle=-90] {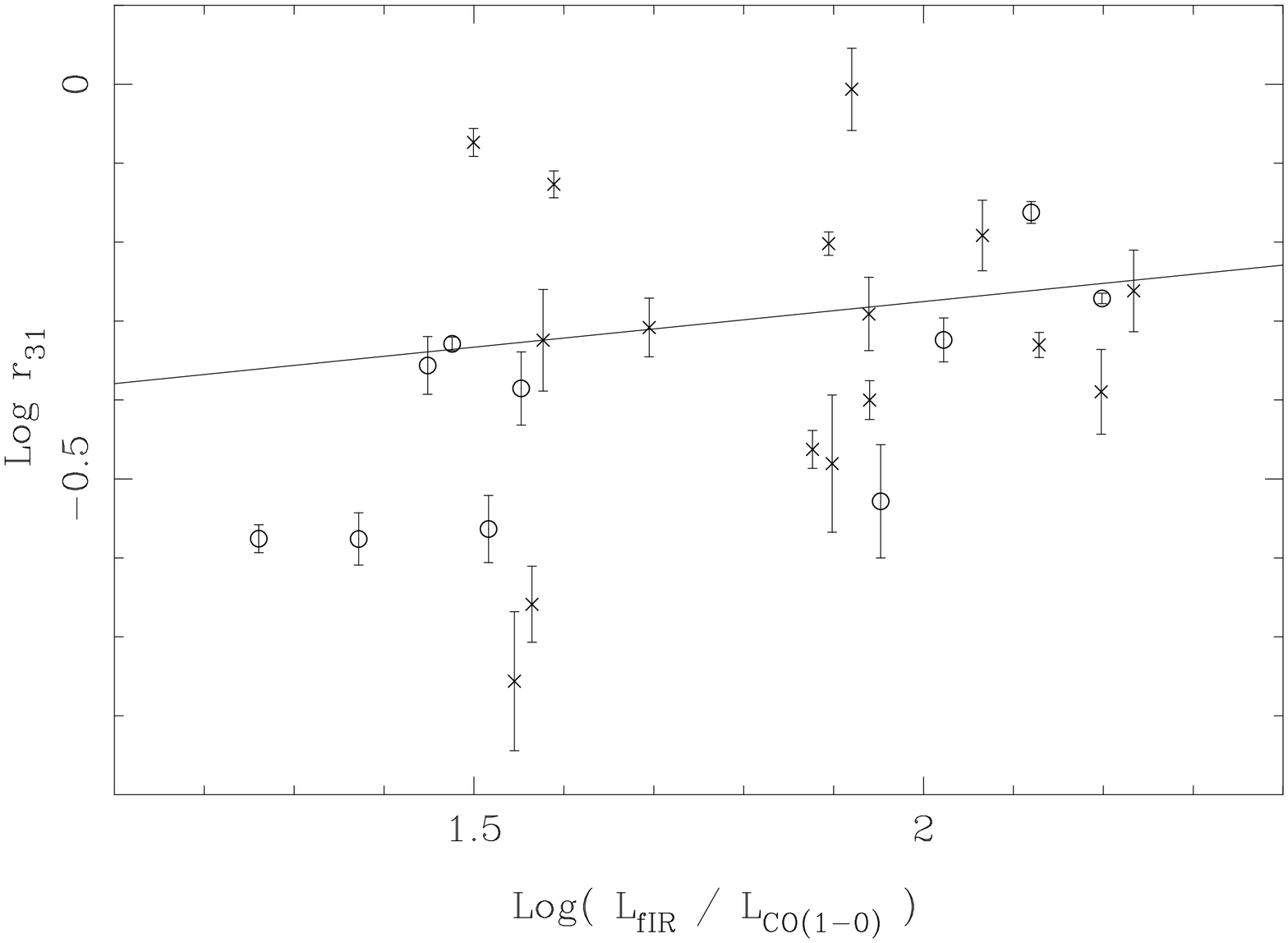}
\caption{A plot of $\log(L_{\rm{fIR}} / L_{\rm{CO(1-0)}} )$ vs.
$\log(r_{31})$ for detected objects in the LIG sample, with symbols defined
as in Figure \ref{log_sep_vs_log_rat}. The best line fit to the data, 
weighted by the random errors, is given and the  Pearson correlation
coefficient is 0.38.}
\label{log_L_fIR_over_LCO10_vs_log_rat} 
\end{center}
\end{figure}

\begin{figure}\begin{center}
\includegraphics [width=2.2in,angle=-90] {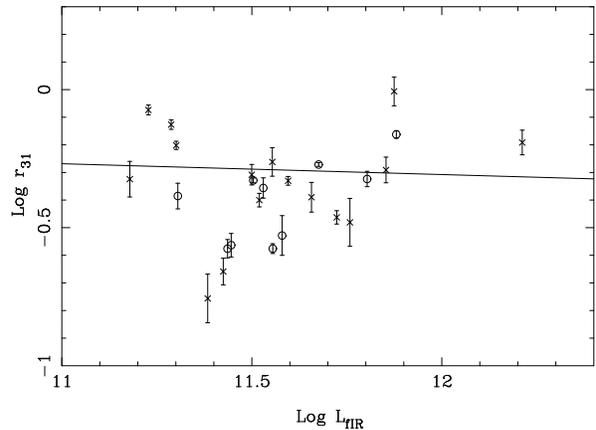}
\caption{ A plot of $\log(r_{31})$ vs. $\log(L_{\rm{fIR}})$ for the
  detected objects in the LIG sample, with symbols defined
as in Figure \ref{log_sep_vs_log_rat}. The best
  line fit to the data, weighted by the random errors, is given and
  the Pearson correlation coefficient is 0.21. }
\label{log_fIR_vs_log_rat} 
\end{center}
\end{figure}

  A common quantity used to determine the star formation efficiency of
  a galaxy or galactic region, is $\log(L_{\rm{fIR}} /
  L_{\rm{CO(1-0)}} )$, where we assume here that the far-infrared
  luminosity is a result of emission from dust heated by UV-photons
  emitted from star formation rather than AGN.  Shown in Figure
  \ref{log_L_fIR_over_LCO10_vs_log_rat} is a plot of $\log(r_{31})$
  vs.  $\log(L_{\rm{fIR}} / L_{\rm{CO(1-0)}} )$. A weak correlation
  ($r=0.38$) is seen, suggesting that there is an increase in gas
  excitation in galaxies that show an increased SFE. This is not
  surprising, as one might expect to observe higher gas excitation in
  regions where star formation is proceeding vigorously. No significant
  correlation is seen between $\log(r_{31})$ and $\log(L_{\rm{fIR}})$
  (Figure \ref{log_fIR_vs_log_rat}, $r=0.21$).

 Our observations and analysis hint at an increase in excitation, as
 traced by CO(3-2), with the temporal progression of the merger. The
 uncertainties within our current data along with the relatively small
 number of sources per unit change in $L_{fIR}$ and nuclear separation
 make it difficult to state this with a sufficient degree of
 confidence, however. For the objects larger than a beamwidth,
 wide--angle jiggle-mapping with the newly commissioned HARP focal
 plane array \citep{2003SPIE.4855..338S} will enable the full extent
 of the CO(3-2) emission from the merging galaxies to be captured more
 accurately, enabling a more accurate line ratio to be determined.

\subsection{Contribution to the 850 $\mu$m continuum flux} 

  An 850 $\mu$m / 450 $\mu$m continuum emission survey has recently
  been completed \citep{inprep:2009} for the galaxies in the Gao and
  Solomon sample in order to map the the distribution of the dust
  emission and determine dust masses and temperatures. This has been
  made with SCUBA, a bolometer array receiver on the JCMT
  \citep{1999MNRAS.303..659H}. The SCUBA 850~$\mu$m camera has a total
  bandwidth of $\sim$30 GHz set by a bandpass filter with a centre
  frequency of 347~GHz. If the redshift of an object is such that the
  CO(3-2) line falls within the bandpass of this filter, the CO(3-2)
  line emission will contaminate the continuum emission from the dust
  \citep{2000ApJ...537..631P}. The CO(3-2) observations made here are
  thus very useful in determining the contribution of the CO(3-2) line
  emission to the 850~$\mu$m flux continuum flux measured using SCUBA.
 
  The CO(3-2) contribution to the measured SCUBA  850~$\mu$m flux can
 be calculated  using,

\begin{equation}
S(V) = {{2k\nu_L^2}\over{c^2}} \Omega_b {{g(V)}\over {\int g(V) dV}}
I_{\rm{CO(3-2)}} \, [\rm{W m^{-2} Hz^{-1} beam^{-1}}]
\label{scuba_cont}
\end{equation} 

\noindent \citep{2004MNRAS.349.1428S} where $\nu_L$ is the line
frequency, $\Omega_b$ is the main-beam solid angle and
$I_{\rm{CO(3-2)}} = \int T_{\rm{mb}} \, dV$ is the velocity integrated
main-beam brightness temperature in $\rm{K\, km\, s^{-1}}$. For
$g(V)$, the bandpass profile of the filter expressed as a function of
the recession velocity $V=cz$, we have adopted the same dual Gaussian
form used by the SLUGS survey in \citet{2004MNRAS.349.1428S}.

 Table \ref{scuba_tab} shows the CO(3-2) contribution to the measured
 SCUBA 850~$\mu$m flux for the observed objects in the sample.  As the
 SCUBA 850~$\mu$m fluxes are measured for the sample, these data will
 enable the correction for the CO(3-2) contamination of the dust
 emission to be made. These corrected 850~$\mu$m fluxes will be
 published in a forthcoming paper -- preliminary results show that the
 mean CO(3-2) contribution for the sample is 22\% and the median
 contribution is 17\% \citep[][priv. comm.]{priv:2009}. An 850~$\mu$m
 map of Arp~236 has been made \citep{1999AJ....118..139F} and the
 total flux at 850~$\mu$m is reported at 273 mJy. The CO(3-2)
 contribution to this flux will be around 91.4 mJy (33\%). These
 results demonstrate the importance of determining the correction both
 when fitting dust emission models to observed fluxes to determine
 dust temperatures, and particularly when determining total dust
 masses, which are usually taken to be proportional to the 850~$\mu$m
 flux \citep{2000MNRAS.315..115D}.

\begin{table}
\begin{center}
\begin{spacing}{1}
\begin{threeparttable}
\centering
\scriptsize
\begin{tabular}{lccc} \hline\hline
Object        &  CO(3-2) contribution       \\
              &  ($\rm{mJy}$) \\ 
              &  (Filter corrected)          \\ \hline
00057+4021    &  2.7      \\      
Arp 236       &  91.4    \\
01077-1707    &  14.9     \\
III Zw 35     &  12.6     \\
Mrk 1027      &  43.7     \\
02483+4302    &  0.6      \\
UGC 2369      &  13.8     \\
03359+1523    &  7.4      \\
04232+1436    &  0.0      \\
Arp 55        &  21.8     \\
10039-3338    & 9.6      \\
10190+1322    & 0.0       \\
10565+2448    & 9.9      \\
Arp 299       & 103.1     \\
13001-2339    & 27.2      \\%
Arp 238       & 11.0      \\
NGC 5256      & 27.1      \\
14348-1447    & 0.0       \\
Arp 302       & 37.4      \\
Mrk 848       & 9.0       \\
Mrk 673       & 10.7      \\
NGC 6090      & 38.7      \\
17132+5313    & 1.6       \\
NGC 6670      & 30.6      \\
19297-0406    & 0.0       \\
20010-2352    & 2.2       \\
II Zw 96      & 15.6      \\
NGC 7592      & $<$4.1    \\ \hline
\end{tabular}
\end{threeparttable}
\caption{Contributions of the  CO(3-2) line to the 850~$\mu$m SCUBA
  continuum flux. }
\label{scuba_tab}
\end{spacing}
\end{center}
\end{table}

\section{Conclusions and further work}
\label{conclusions}

  We have observed the CO(3-2) emission from a sample of IR-selected
 merging galaxies using the single-pixel 350 GHz receiver at the
 JCMT. We have found that the CO(3-2) to CO(1-0) line ratio,
 $r_{31}$, is less than unity for all but one object in our merging
 sample with a sample average of $r_{31}=0.47$. LVG modelling
 indicates that this is consistent with $T_k$=(30--50) K, $n=10^3
 \rm{cm}^{-3}$ and $\Lambda \equiv [\rm{CO/H_2}]/(dV/dR) =10^{-5}$
 $(\rm{km \, s^{-1}}\, \rm{pc}^{-1})^{-1} $, although warmer ($T_k
 \sim$ 80--110~K) and more diffuse $(n \sim 3 \times 10^2
 \rm{cm}^{-3})$ conditions for the gas phase are not ruled out.  We
 have calculated the CO(3-2) contribution to the 850~$\mu$m
 continuum flux for the galaxies in our sample. We have found that
 this contribution can be a large fraction of the 850~$\mu$m flux
 underlining the importance of making CO(3-2)-based flux corrections
 when deriving SED fits or dust masses from 850~$\mu$m fluxes. Our
 correction will be presented in a forthcoming paper presenting the
 850~$\mu$m SCUBA data for this galaxy set.

  The spread of line ratios observed in our merging sample was
 broadly similar to that measured in both an IR-selected subsample
 of the SCUBA Local Universe Galaxy Survey (SLUGS)
 \citep{2003ApJ...588..771Y}, and a sample of normal galaxies
 \citep{1999A&A...341..256M}. We observe a weak anti-correlation
 between $\log(r_{31})$ and $\log(\rm{sep; kpc})$ in our sample, a
 result consistent with gas excitation increasing as the merger
 progresses. A plot of line ratio versus star formation efficiency
 ($\log(r_{31})$ vs.  $\log(L_{\rm{fIR}} / L_{\rm{CO(1-0)}} )$)
 also shows a weak correlation, as gas excitation increases with
 increased star formation efficiencies. These correlations are
 suggestive of an increase in the excitation of the molecular gas,
 as measured by the CO(3-2)/CO(1-0) line ratio, with the
 progression of galactic mergers. However, studies of line ratios
 of these kind are often unavoidably affected by uncertainties in
 line luminosities arising both from random errors and systematic
 errors in the absolute flux calibration, which can be as high as
 20\%. Uncertainties also arise from assumptions that often must be
 made about the angular size and shape of the CO(3-2) emission
 regions relative to the telescope's beam. These latter
 uncertainties will hopefully be reduced as more small sources
 ($D_A \sim$ a few arcseconds) are mapped interferometrically, and
 as more extended sources ($D_A \gtrsim 10''$) are mapped with the
 next generation of multi-element focal-plane heterodyne arrays,
 such as HARP-B on the JCMT.

  Our future work includes an LVG analysis in which we will combine
 our CO(3-2) data with HCN line measurements for the same galaxy
 set to systematically probe the gas properties across the merging
 sequence. We also intend improve our total luminosity estimates
 for our more extended sources by mapping them with the newly
 commissioned HARP-B focal-plane array.

\appendix

\section{Systematics and error analysis}
\label{errors}
 As well as the systematic calibration errors of around $\pm$20\%,
there are random errors arising from noise across each spectral
channel. This noise leads to errors in determining the integrated
spectral line and in determining the overall subtracted baseline
level. Following \citet{Gao:phd}, these random errors were combined
using

\begin{equation}
\sigma I= \sqrt{\sigma I^2_{\rm{line}}+\sigma I^2_{\rm{base}}}=\sigma
T_{\rm{rms}} \sqrt{\Delta W \Delta V}/\sqrt{1-{\Delta W}/W}.
\label{err3}
\end{equation}

\noindent where $\sigma T_{\rm{rms}}$ is the RMS noise on a single velocity
channel of width $\Delta V$, $\Delta W$ is the full width to zero
intensity of the line and $W$ is the total velocity coverage of the
observation. These random errors in the velocity integrated main-beam
brightness temperature, $\sigma I$, are presented separately from any
overall calibration errors.

\section{Deriving the global CO(3-2) luminosities from multiple pointings}
\label{multiple}

\begin{table*}
\caption{Details of the combination of the observations for the
  galaxies observed with two pointings. For each galaxy the model with
  a small emitting region (1$''$) was chosen, and the value of
  $I_{\rm{TOTAL}}$ marked with an asterix used in calculating the
  final CO(3-2) luminosity.}
\begin{center}
\scriptsize
\begin{tabular}{lcccccccccccccc} \hline\hline
Object  & Pointing 1 & $\alpha_1$ & $D_{0,1}$ & $\rm{PA}_1$ & Pointing 2 & $\alpha_2$ & $D_{0,2}$ & $\rm{PA}_2$ & $I_1$ & $I_2$ &$I_{01}$ & $I_{02}$ & $I_{\rm{TOTAL}}$ \\ \hline 

Arp 55 &  (0,0)      & 0 &   1    &     0      &  (+10,+7) &  0    &   1    &   0    &   28.1 & 18.3  & 26.0 & 14.2 & 40.2& \\
Arp 55 &  (0,0)      & 0 &   7    &     0      &  (+10,+7) &  0    &   7    &   0    &   28.1 & 18.3  & 30.8 & 15.5 & 46.3& \\
Arp 55 (corrected) &  (0,0)      & 0 &   1    &     0      &  (+10,+7) &  0    &   1    &   0    &   28.1 & 18.3  & 24.4 & 20.7 & 45.1*& \\\hline

NGC 5256 &  (0,0)      & 0 &   1    &     0      &  (-7,-9)  &  0    &   1    &   0   &   18.4 & 29.0 & 13.1  & 26.5 & 39.6*& \\      
NGC 5256 &  (0,0)      & 0 &   7    &     0      &  (-7,-9)  &  0    &   7    &   0   &   18.4 & 29.0 & 13.9  & 31.6 & 45.5 & \\ \hline

II Zw 96&  (0,0)      & 0 &   1    &     0      &  (+8,-9)  &  0    &   1    &   0   &   13.7 & 17.9 & 11.0  & 16.2 & 27.2* & \\      
II Zw 96 &  (0,0)      & 0 &   7    &     0      &  (+8,-9)  &  0    &   7    &   0   &  13.7 & 17.9 & 12.3  & 19.0 & 31.3 & \\ \hline
\end{tabular}
\end{center}
\label{gauss_model_table}
\end{table*}

 For galaxies with a projected nuclear separation $\sim 14 ''$ the
 procedure followed for combining the two measured intensities was
 similar to that used by \citet{1999AJ....118..145Z}. Each galaxy was
 modelled as a Gaussian brightness distribution of diameter $D_{0}$
 and inclination $\alpha$ with a position angle PA. This procedure
 allowed the intrinsic intensities for each pointing $I_{01}$ and
 $I_{02}$ to be derived from the measured intensity for each pointing
 $I_{1}$ and $I_{2}$. For II~Zw~96, NGC~5256 and Arp~55 detections
 were observed in both pointings and the parameters used in the source
 models are shown in Table \ref{gauss_model_table}. Objects where a line was
 only detected in one of two pointings were treated identically to the
 objects for which a single pointing was made, with the assumption
 that the CO(3-2) emitting region is small.

 The optical images of Arp~55, NGC~5256 and II~Zw~96 (Figure
 \ref{spec1} (w),(q) and (t)) suggest collisions between two face-on
 galaxies, and the optical brightness distributions are near circular
 so values of $\alpha=0$ and PA = 0 were chosen for both components.

 For each object, the intrinsic CO(3-2) intensities $I_{01}$, $I_{02}$
 were calculated for two cases: (i) assuming that the half power width
 of the CO(3-2) emitting regions were both small compared to the 14''
 beam ($D_0=1''$), and (ii) assuming the half power width of the
 CO(3-2) emitting regions were around one half of a beam size
 ($D_0=7''$).  This gave an assessment of the sensitivity of the model
 to the unknown beam filling factors. The results for the two
 differing source sizes are shown in Table
 \ref{gauss_model_table}. For calculating the final combined CO(3-2)
 intensity, the first model (i) corresponding to two small emitting
 regions was chosen in keeping with the assumptions made for the
 objects observed with a single pointing. Note that for Arp 55, it was
 found that the galaxy was inadvertently observed with a map centre of
 09:15:54.9~+44:19:54.4~(J2000) rather than the required co-ordinate
 of 09:15:54.7~ +44:19:51 (see Figure \ref{spec1}(w)). We thus
 incorporated this additional offset of $(+1.9'',+3.4'')$ in each of
 the two pointings into our model, leading to an estimated total flux
 of 45.1~K km/s (see the ``Arp 55 (corrected)'' row in Table
 \ref{gauss_model_table} above).

\section{CO(3-2) spectra and pointings}

 The main beam efficiency-corrected spectra together with the pointing
 centres superimposed on a Digital Sky Survey optical image, are shown
 in Figure \ref{spec1}.

\section*{Acknowledgments}

 We would like to thank all of the support staff at the JCMT, and the
 visiting observers who took data for this project when the weather
 was too poor for their own projects. Y.~Gao's research is partly
 supported by grants \#10833006 \& \#10621303 of China NSF.

\bibliography{lig_paper_rev2a}

\bsp

\begin{figure*}
\begin{center}
\includegraphics [width=6.0in, height=9.0in,angle=0] {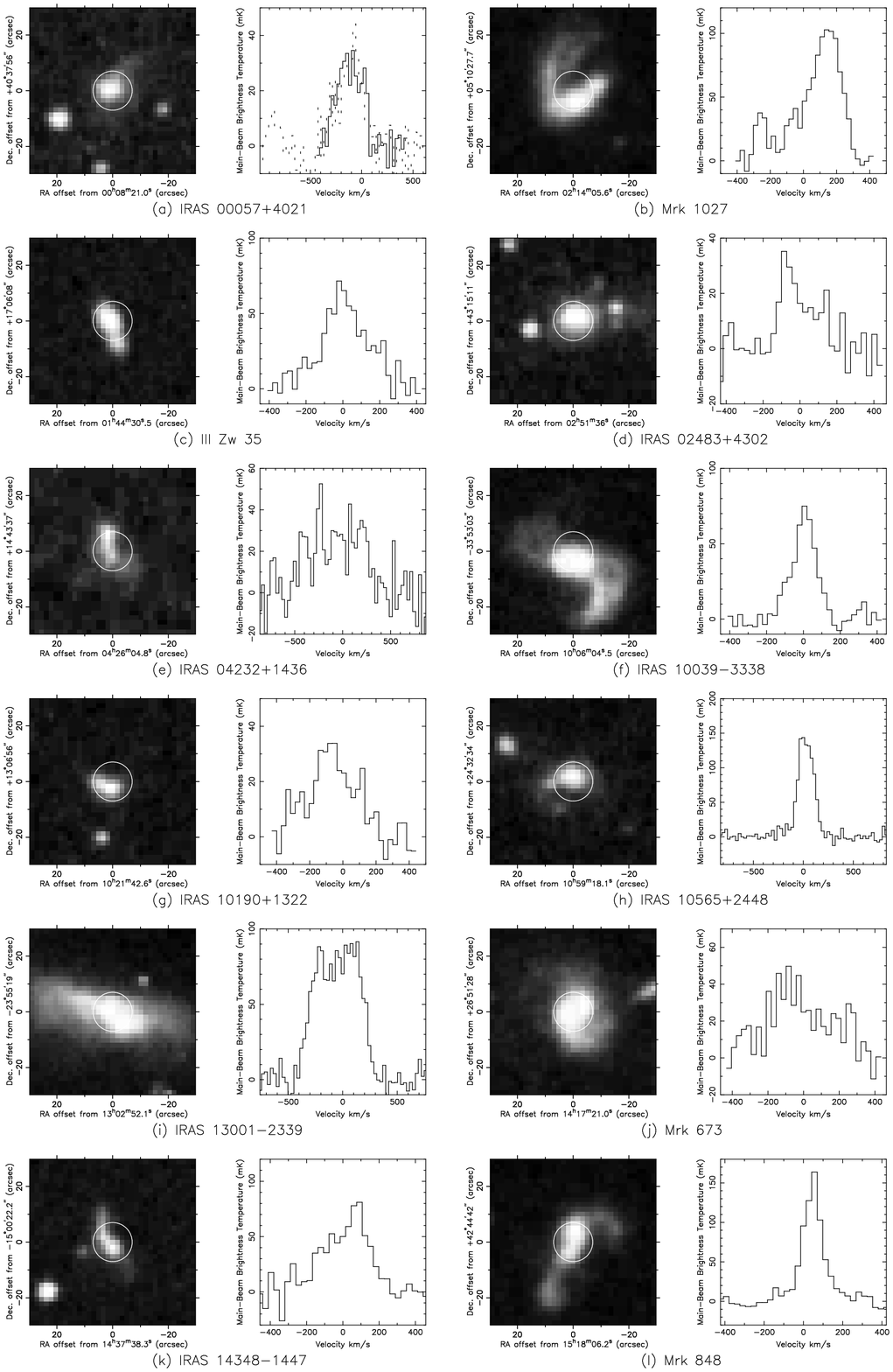}
\end{center}
\caption[]{CO(3-2) spectra and optical images (with pointings and
  beam size). The optical image wavelength is 645 nm (band 103aE)
  (taken from the STScI Digitized Sky Survey (DSS)
  \citep{1994Ap&SS.217...27K}, {\tt
    http://archive.stsci.edu/cgi-bin/dss\_form} ). Where the object
  was observed with both narrow (900~MHz) and wide (1800~MHz)
  spectrometer bandwidths, both spectra are show, with the wideband spectrum shown as a dotted line.}
\label{spec1}
\end{figure*}

\begin{figure*}
\begin{center}
\includegraphics [width=6.0in, height=3.0in,angle=0] {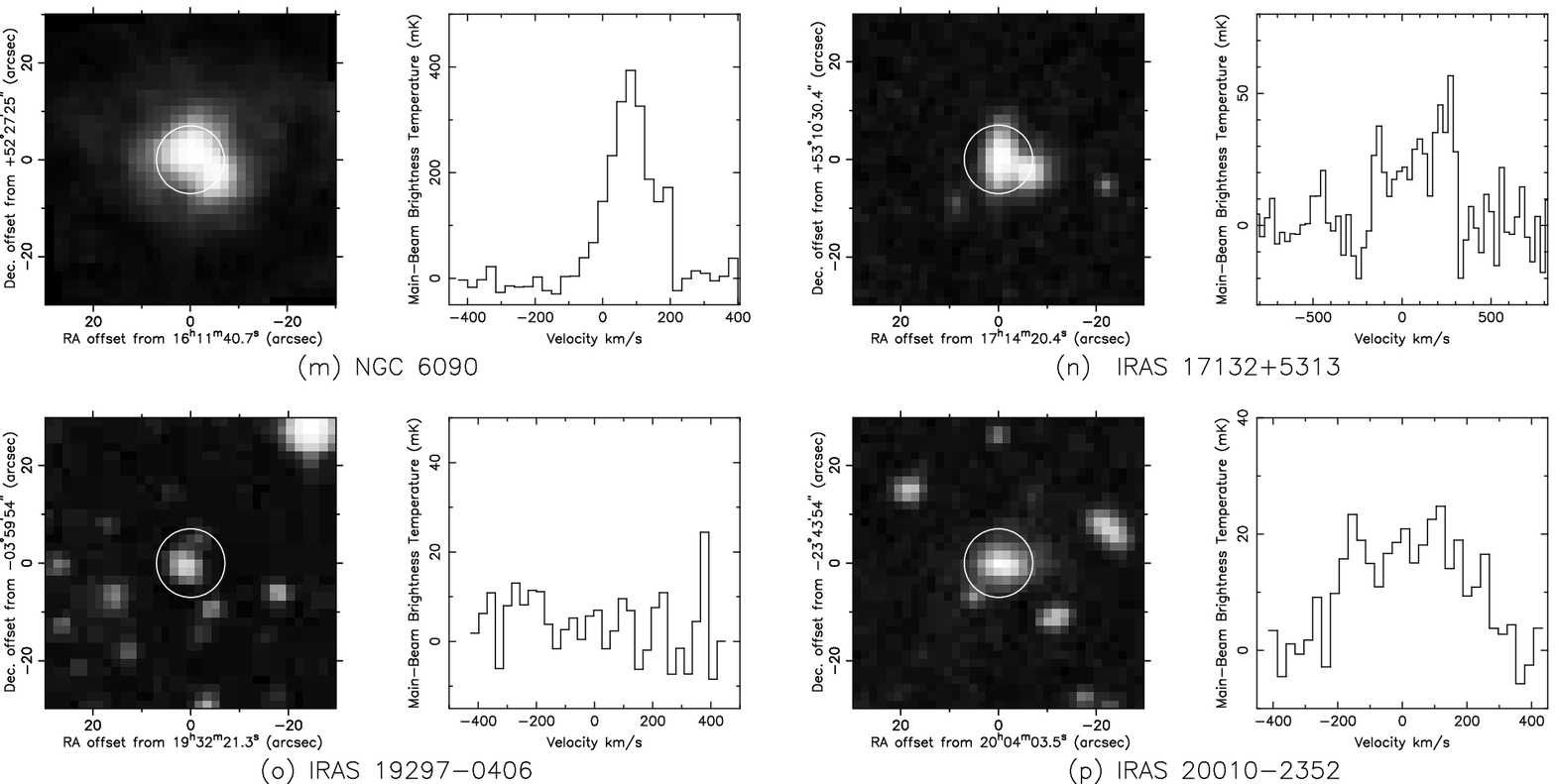}
\end{center}
\contcaption{}
\end{figure*}

\begin{figure*}
\begin{center}
\includegraphics [width=4.5in, height=9.0in,angle=0] {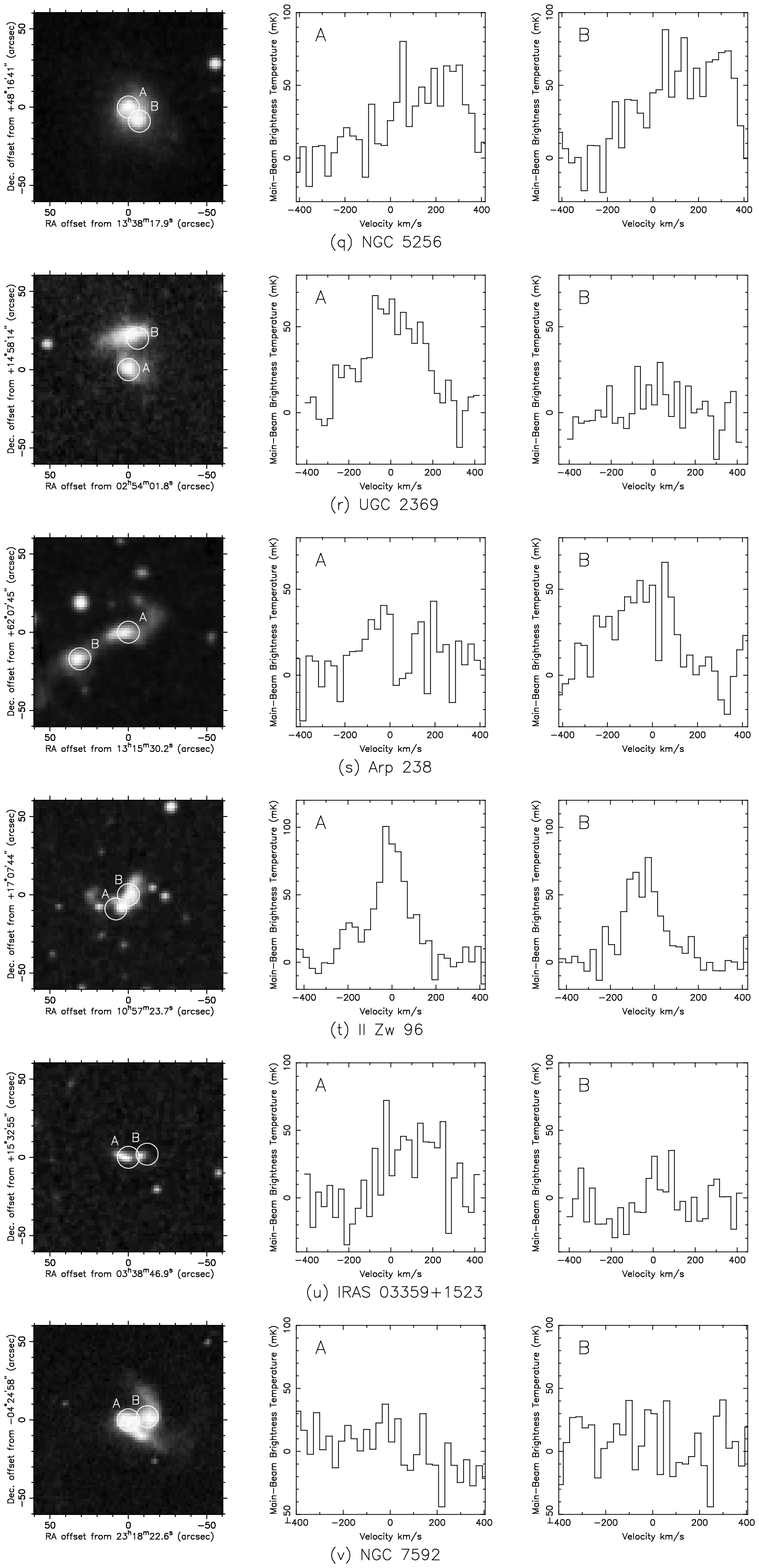}
\end{center}
\contcaption{}
\end{figure*}

\begin{figure*}
\begin{center}
\includegraphics [width=4.5in, height=1.5in,angle=0] {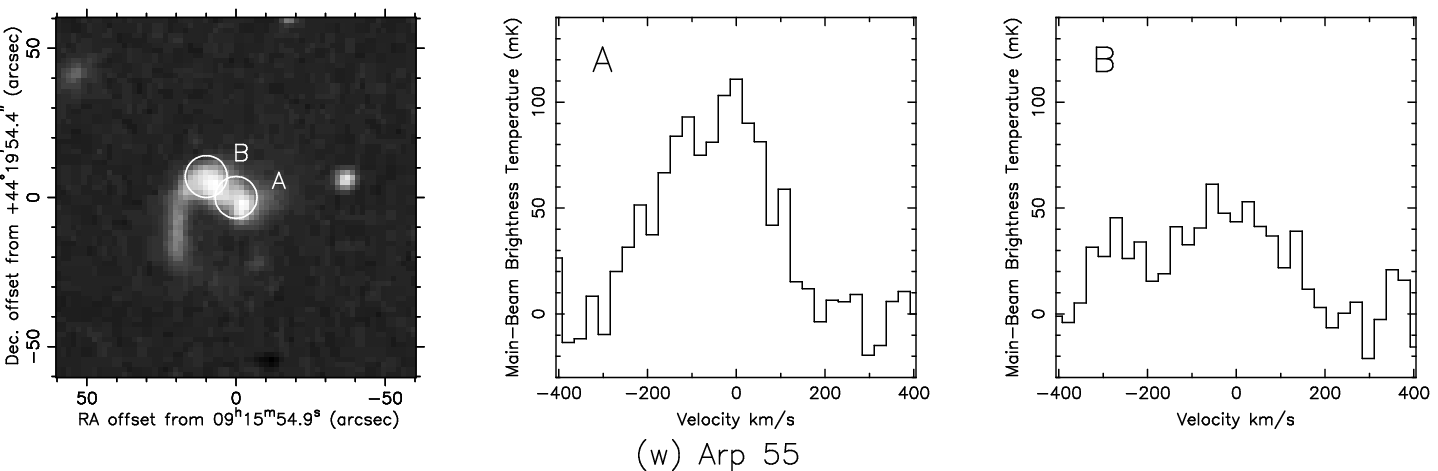}
\end{center}
\contcaption{}
\end{figure*}

\begin{figure*}
\begin{center}
\includegraphics [width=6.0in, height=9.0in,angle=0] {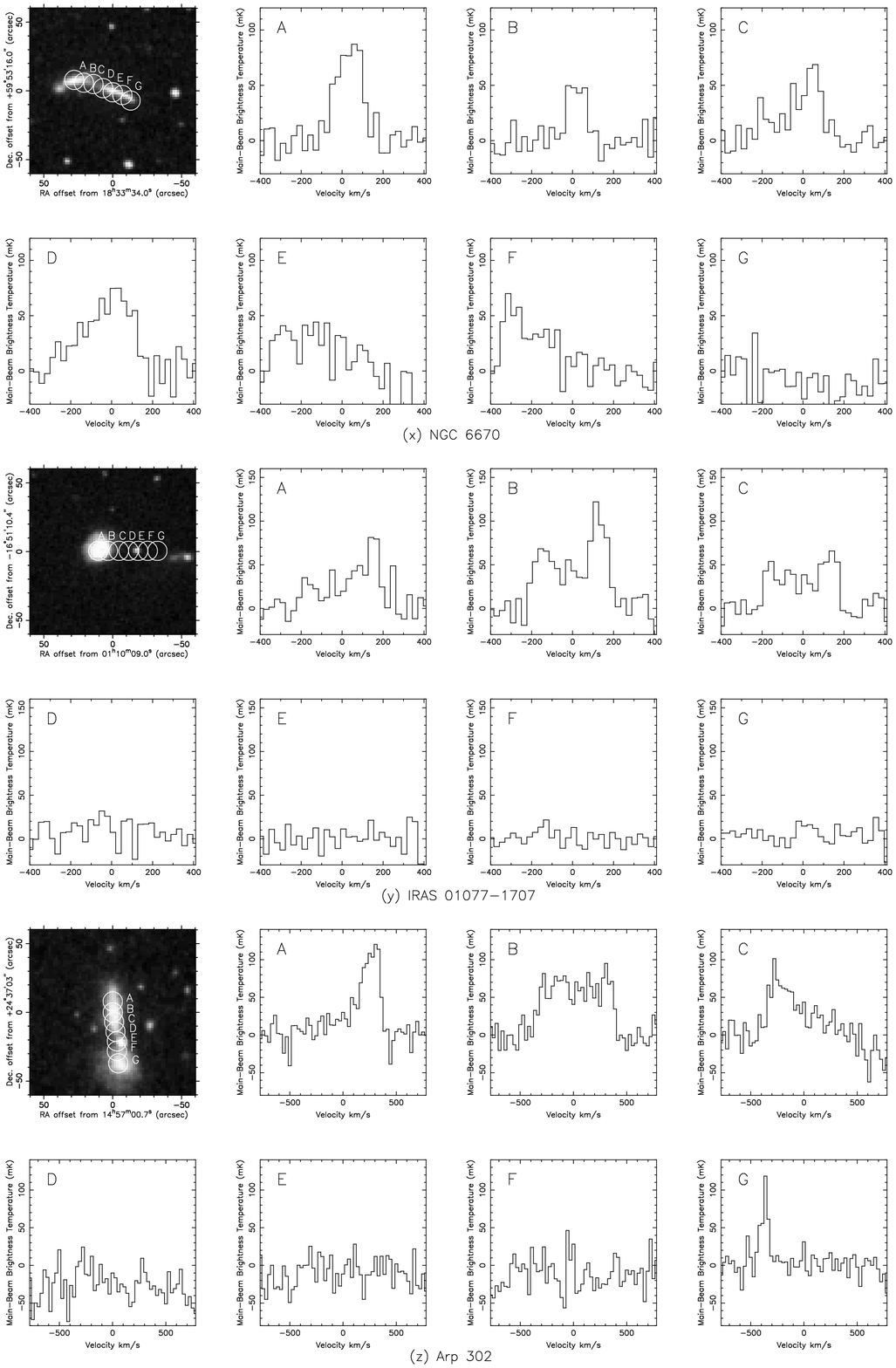}
\end{center}
\contcaption{}
\end{figure*}

\begin{figure*}
\begin{center}
\includegraphics [width=6.0in, height=1.5in,angle=0] {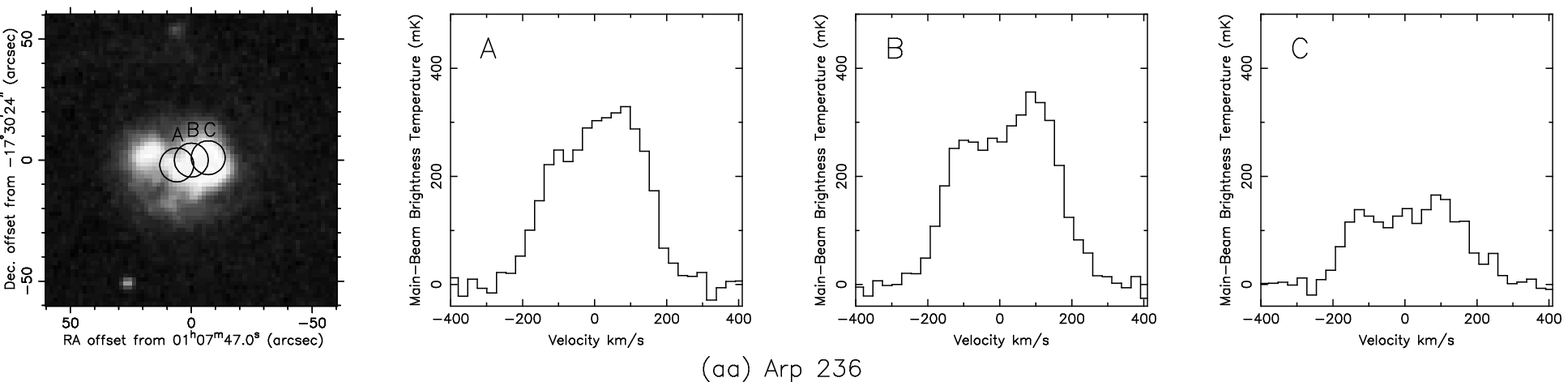}
\end{center}
\contcaption{}
\end{figure*}

\begin{figure*}
\begin{center}
\includegraphics [width=6.0in, height=3.0in,angle=0] {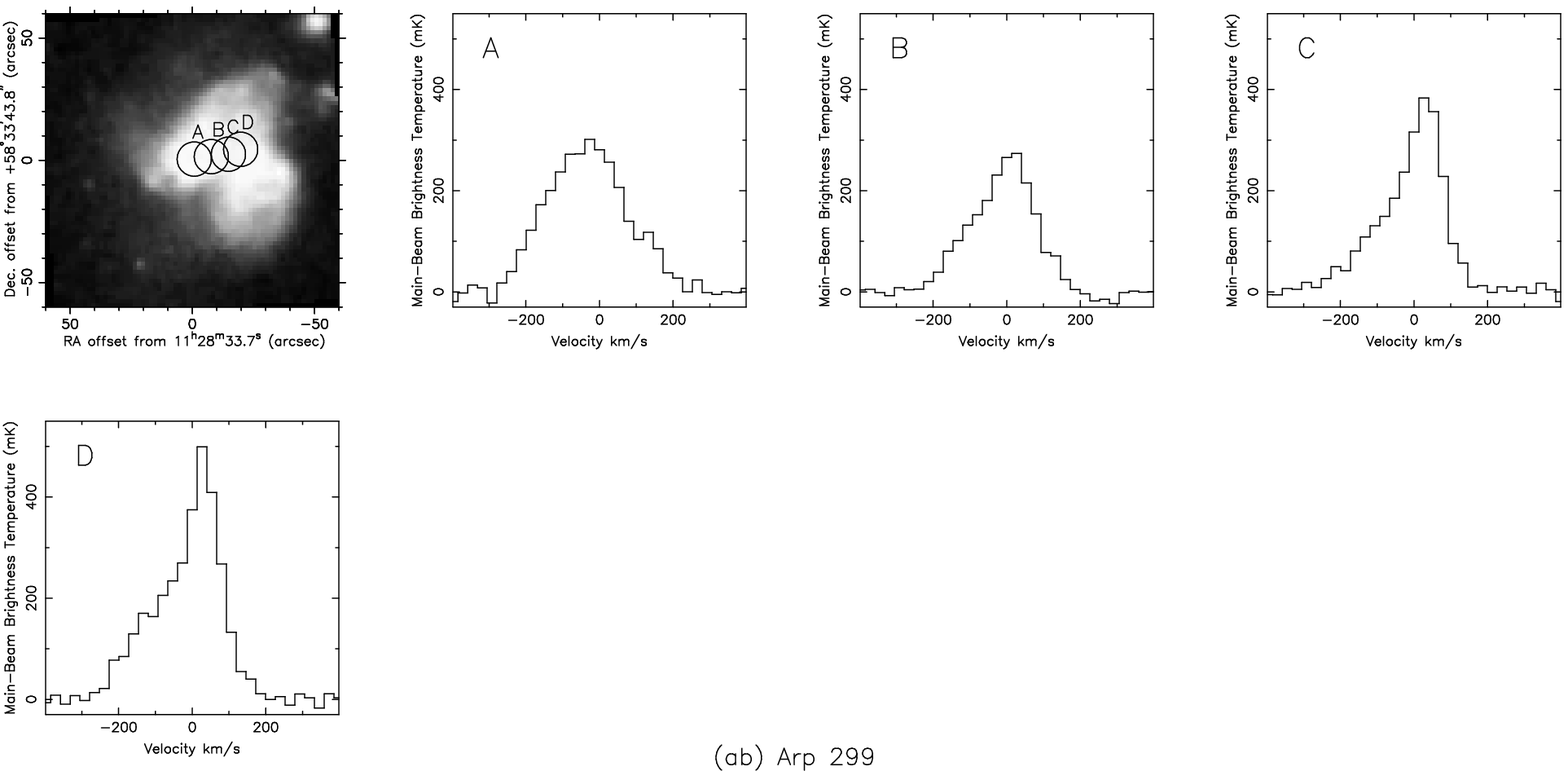}
\end{center}
\contcaption{}
\end{figure*}

\label{lastpage}

\end{document}